\documentclass[9pt]{article}
\usepackage[T1]{fontenc}
\usepackage{lmodern}
\usepackage[utf8]{inputenc}      
\usepackage{amsmath, amssymb, amsfonts} 
\usepackage{graphicx}            
\usepackage{authblk}             
\usepackage{hyperref}            
\usepackage{cite}                
\usepackage{geometry}            
\usepackage{multirow}
\usepackage{wrapfig}
\usepackage{xspace}
\usepackage{hyphenat}

\newcommand{\Korea}{Department of Physics and Astronomy, Seoul National University, Seoul 151-747, Korea}
\newcommand{\Rochester}{Department of Physics and Astronomy, University of Rochester, Rochester, NY  14627, USA}

\newcommand{\seff}{\ensuremath{\sin^2\theta^\ell_{\mathrm{eff}}}\xspace}
\newcommand{\AFB}{\ensuremath{A_{\rm FB}}\xspace}
\newcommand{\Wasym}{\ensuremath{W_{\rm asym}}\xspace}

\title{Summary of the Precision Measurements of the Electroweak Mixing Angle in the Region of the Z pole}

\author[1]{Arie~Bodek}
\author[1]{Hyon-San~Seo}
\author[2]{Un-Ki~Yang}

\affil[1]{\Rochester}
\affil[2]{\Korea}

\date{\normalsize \textit{Presented by Hyon-San Seo at the 32nd International Symposium on Lepton Photon Interactions at High Energies,\\ Madison, Wisconsin, USA, August 25--29, 2025}}

\begin{document}
\maketitle

\begin{abstract}
This contribution presents an overview of an improved extraction of the effective leptonic weak mixing angle, \seff, based on the published CMS measurement of the forward–backward asymmetry in Drell–Yan events at 13 TeV~\cite{Bodek:2025kxl}. While the original CMS analysis~\cite{CMS:2024ony} achieved a significant reduction in experimental uncertainties, its overall precision remains limited by residual uncertainties in the parton distribution functions (PDFs). This proceeding highlights the impact of incorporating complementary CMS measurements that probe different combinations of parton densities, thereby providing additional PDF constraints beyond those obtained from the asymmetry measurement alone.
The improved analysis leads to a substantially reduced total uncertainty, yielding \seff = 0.23156$\pm$0.00024. This result is consistent with the Standard Model prediction and represents the most precise single determination of this parameter to date.
\end{abstract}

\section{Introduction}\label{sec_newsw2}
The effective leptonic weak mixing angle (\seff) is a key parameter in precision tests of the Standard Model (SM). A recent CMS analysis of the forward-backward asymmetry (\AFB) in Drell-Yan events at 13 TeV~\cite{CMS:2024ony} achieved a landmark precision, comparable to results from electron-positron LEP/SLD colliders. Despite the high experimental precision achieved by the CMS analysis, uncertainties related to the proton's parton distribution functions (PDFs) remain the dominant limitation.

This work summarizes the results obtained using an extended PDF profiling strategy applied to the CMS measurement. We incorporate additional, powerful constraints from CMS measurements of the $W$-boson decay lepton asymmetry~\cite{CMS:2020cph} and $W/Z$ cross-section ratios~\cite{CMS:2024myi}. After establishing consistency with the published CMS measurement, the extended fit strategy leads to a noticeable reduction in the PDF-driven component of the total uncertainty. This approach results in the most precise single measurement of \seff to date.

\section{Reproduction of the CMS 13 TeV analysis}
The CMS extraction of the weak mixing angle is based on a global fit framework that incorporates correlated experimental and theoretical uncertainties~\cite{CMS:2024ony}.
\begin{equation}
 \chi^2(\beta_{\rm exp},\beta_{\rm th}) = 
 \sum_i\frac{(\sigma_i^{\rm exp}+\sum_j\Gamma_{ij}^{\rm exp}\beta_{j,{\rm exp}}-\sigma_i^{\rm th}-\sum_k\Gamma_{ik}^{\rm th}\beta_{k,{\rm th}})^2}{\Delta_i^2} 
 +\sum_j\beta^2_{j, {\rm exp}} + \sum_k\beta^2_{k,{\rm th}}.
\end{equation}

 \begin{figure}[t]
 \centering
\includegraphics[width=5in,height=3.23in]{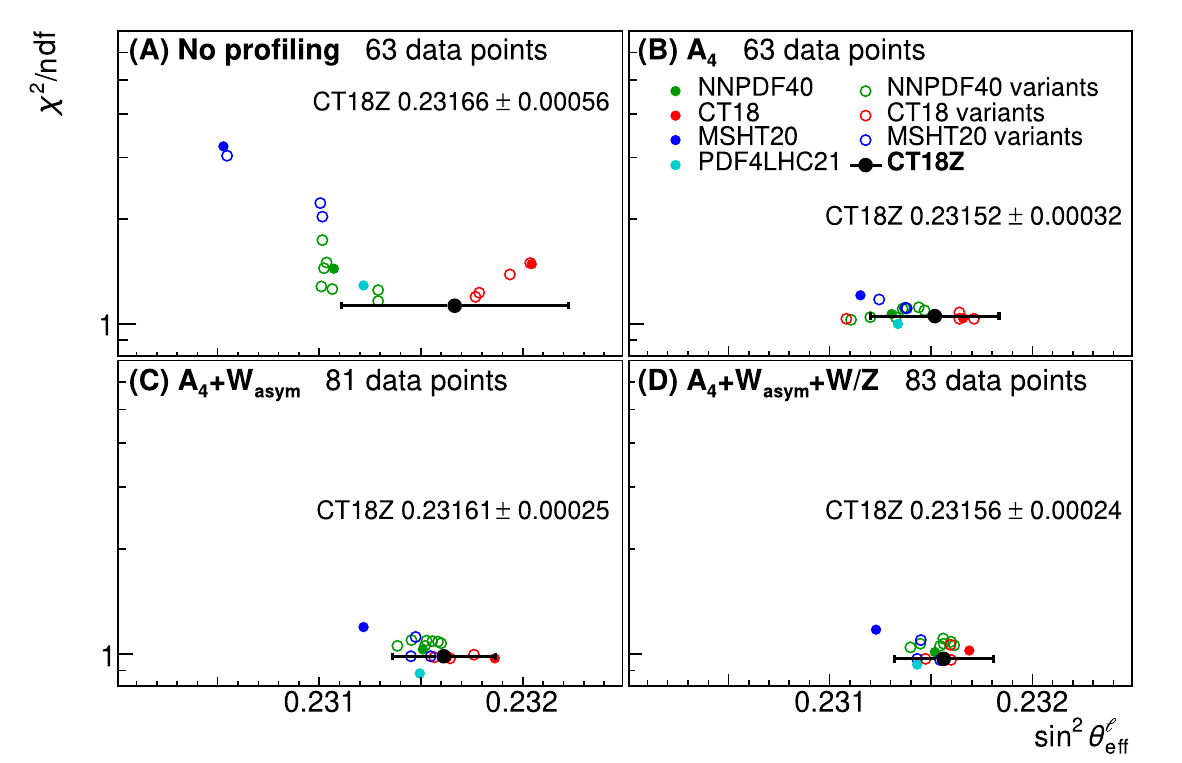}
\caption{Extracted values of \seff from the 13 TeV CMS $A_4$ data  for 19 different PDF sets on the horizontal axis. (A) Before profiling. (B) After profiling with $A_4$.  (C) After profiling with $A_4$ plus $W$ decay lepton asymmetry.   (D) After profiling with $A_4$ plus $W$ decay lepton asymmetry plus W/Z cross section ratios. The vertical axis shows the $\chi^2$ values of the fits divided by the number of degrees of freedom ($N_{\rm data}-1$), where the one degree of freedom corresponds to the free parameter \seff.
}
\label{Fig_2}
\end{figure} 

Correlated experimental and theoretical uncertainties are implemented by introducing nuisance parameters, $\beta_{\rm exp}$ and $\beta_{\rm th}$, which propagate their effects to the measured spectra and theoretical predictions through the corresponding response matrices $\Gamma^{\rm exp}$ and $\Gamma^{\rm th}$.
Within this framework, the index $i$ labels the $N_{\rm data}$ points of the $(|y|, M)$ double-differential $A_4$ dataset, corresponding to 63 entries used in the $A_4$ profiling.
The indices $j$ and $k$ index the nuisance parameters associated with experimental and theoretical uncertainty sources, respectively.
The measured values and their uncorrelated experimental uncertainties are denoted by $\sigma_i^{\rm exp}$ and $\Delta_i$, while the theoretical predictions are given by $\sigma_i^{\rm th}$.
Correlations among experimental uncertainties in the double-differential $A_4$ measurement are accounted for by the covariance matrix $\Gamma^{\rm exp}$.

Modern QCD calculations are used to model the Drell–Yan process, following the setup of the original CMS analysis~\cite{Monni:2020nks,Golonka:2005pn,Chiesa:2019nqb,Chiesa:2024qzd,Barze:2013fru}. And the $Z$-boson transverse-momentum spectrum is reweighted to data, following the original CMS analysis.

The theoretical uncertainty accounts for effects arising from missing higher-order contributions in both QCD and electroweak (EW) calculations.
Electroweak corrections are evaluated using \textsc{POWHEG-Z\_ew}, while PDF-related uncertainties are derived from NLO grids produced with \textsc{MadGraph5-aMC@nlo}~\cite{Alwall:2014hca} and interfaced through \textsc{PineAPPL}~\cite{schwan2024nnpdf, Carrazza:2020gss}.
These components are incorporated into the matrix $\Gamma^{\rm th}$ via nuisance parameters that represent missing higher-order EW effects, PDF Hessian variations, and the effective weak mixing angle \seff itself, which is treated as a free parameter in the fit.
Uncertainties associated with missing higher-order QCD corrections are assessed by repeating the fit under six independent scale variations, with the largest observed deviation from the nominal result taken as the corresponding uncertainty.


Table~\ref{Table1} summarizes the extracted values of \seff, expressed in units of $10^{-5}$, obtained with the 13~TeV $A_4$ distribution both prior to and following the profiling procedure, for a total of 19 PDF sets.
The majority of these PDFs are computed at next-to-next-to-leading-order (NNLO) accuracy in QCD, while a subset corresponds to approximate next-to-next-to-next-to-leading-order calculations (an3lo).
PDF ensembles marked as ``qed'' incorporate Quantum Electrodynamics effects and explicitly include photon parton densities.
Within the NNPDF40 family, the ``mhou'' designation denotes the inclusion of corrections associated with missing higher-order uncertainties.
The MSHT and NNPDF sets adopt asymmetric parameterizations of the strange-quark sea, whereas the CTEQ sets assume a symmetric strange-quark sea, with the exception of the CT18As PDF.

 The  total uncertainties in Table~\ref{Table1}  include contributions from  statistical, experimental systematic, theoretical, and PDF sources.  
The differences in the results obtained with the various PDF sets ({\it before profiling}) largely reflect variations in input datasets, parameterization choices, and flavor assumptions.
If there are sufficient parameters in the PDF sets, then  the profiled  PDFs should be able to describe all data that are used
in the   \seff  profiling analysis with good $\chi^2$.   The columns labeled marked "Diff. from  \textsc{CT18Z}" show the difference between the value of \seff extracted with each of the 19 PDF sets
and that obtained using the nominal \textsc{CT18Znnlo} PDF set. 

For the nominal \textsc{CT18Znnlo} set, the profiled value obtained from the 13 TeV $A_4$ data alone is in very good agreement with the published CMS result (Table~\ref{Table1}), which validates our implementation of the CMS analysis in \textsc{xFitter}~\cite{Alekhin:2014irh,HERAFitterdevelopersTeam:2015cre}.
%

\begin{table}[t]								
\scriptsize
\begin{center}																									
\begin{tabular}{|c||c|c|c||c|c|c||c|c|c|} \hline		
\multirow{4}{*}{PDF}    &\multicolumn{3}{c||}{$A_{4}$ without profiling} &\multicolumn{3}{c||}{$A_{4}$ profiling} &\multicolumn{3}{c|}{$A_{4}+\Wasym+W/Z$ profiling}	\\
    &\multicolumn{3}{c||}{(63 data points)} &\multicolumn{3}{c||}{(63 data points)} &\multicolumn{3}{c|}{(83 data points)}	\\ \cline{2-10}
    & \multirow{3}{*}{\seff} & Diff. & \multirow{3}{*}{$\chi^2$} & \multirow{3}{*}{\seff} & Diff. & \multirow{3}{*}{$\chi^2$} & \multirow{3}{*}{\seff} & Diff. & \multirow{3}{*}{$\chi^2$} \\
	&  & from & & & from & & & from & \\
	&  & \textsc{CT18Z} & & & \textsc{CT18Z} & & & \textsc{CT18Z} & \\ \hline\hline	

NNPDF40	&				&		&		&				&		&		&				&		&		\\	
nnlo\_as\_01180\_hessian~\cite{NNPDF:2021njg}	&	23107	$\pm$	49	&	-59	&	89	&	23130	$\pm$	24	&	-22	&	66	&	23152	$\pm$	23	&	-4	&	83	\\	
nnlo\_as\_01180\_qed~\cite{NNPDF:2024djq}	&	23102	$\pm$	46	&	-64	&	108	&	23144	$\pm$	23	&	-8	&	69	&	23161	$\pm$	22	&	5	&	87	\\	
nnlo\_as\_01180\_mhou~\cite{NNPDF:2024dpb}	&	23101	$\pm$	45	&	-65	&	79	&	23110	$\pm$	27	&	-42	&	64	&	23145	$\pm$	24	&	-11	&	88	\\	
nnlo\_as\_01180\_qed\_mhou~\cite{NNPDF:2024djq}	&	23103	$\pm$	46	&	-63	&	93	&	23136	$\pm$	24	&	-16	&	69	&	23156	$\pm$	23	&	0	&	88	\\	
an3lo\_as\_01180~\cite{NNPDF:2024nan}	&	23129	$\pm$	46	&	-37	&	72	&	23132	$\pm$	25	&	-20	&	65	&	23154	$\pm$	23	&	-2	&	87	\\	
an3lo\_as\_01180\_qed~\cite{NNPDF:2024nan}	&	23129	$\pm$	49	&	-37	&	78	&	23147	$\pm$	23	&	-5	&	68	&	23160	$\pm$	22	&	4	&	89	\\	
an3lo\_as\_01180\_mhou~\cite{NNPDF:2024nan}	&	23106	$\pm$	47	&	-60	&	78	&	23120	$\pm$	26	&	-32	&	65	&	23140	$\pm$	25	&	-16	&	86	\\	
an3lo\_as\_01180\_qed\_mhou~\cite{NNPDF:2024nan}	&	23102	$\pm$	44	&	-64	&	90	&	23137	$\pm$	23	&	-15	&	69	&	23156	$\pm$	23	&	0    &	91	\\	 \hline
 \textsc{CTEQ}          	&				&		&		&				&		&		&				&		&		\\	
 \textsc{CT18nnlo}~\cite{Hou:2019efy}	&	23204	$\pm$	64	&	38	&	92	&	23166	$\pm$	36	&	14	&	64	&	23169	$\pm$	24	&	13	&	84	\\	
{\bf \textsc{CT18Znnlo}}~\cite{Hou:2019efy} 	&	23166	$\pm$	56	&	0	&	70	&	23152	$\pm$	32	&	0	&	65	&	{\bf 23156	$\pm$	24}	&	0	&	79	\\	
\textsc{CT18Annlo}~\cite{Hou:2019efy}  	&	23179	$\pm$	54	&	13	&	76	&	23164	$\pm$	28	&	12	&	67	&	23160	$\pm$	23	&	4	&	79	\\	
 \textsc{CT18Xnnlo}~\cite{Hou:2019efy} 	&	23194	$\pm$	53	&	28	&	86	&	23171	$\pm$	30	&	19	&	64	&	23160	$\pm$	24	&	4	&	87	\\	
\textsc{CT18qed-proton}~\cite{Xie:2023qbn}	&	23204	$\pm$	64	&	38	&	93	&	23164	$\pm$	36	&	12	&	64	&	23155	$\pm$	24	&	-1	&	80	\\	
\textsc{CT18As\_LatNNLO}~\cite{Hou:2022onq}                          	&	23177	$\pm$	75	&   11	&	74	&	23108	$\pm$	43	&	-44	&	64	&	23147	$\pm$	35	&	-9	&	80	\\	\hline
MSHT20             	&				&		&		&				&		&		&				&		&		\\	
MSHT20nnlo\_as118~\cite{Bailey:2020ooq}                         	&	23053	$\pm$	46	&	-113	&	200	&	23115	$\pm$	30	&	-37	&	75	&	23123	$\pm$	26	&	-33	&	96	\\	
MSHT20qed\_nnlo~\cite{Cridge:2021pxm}  	&	23055	$\pm$	59	&	-111	&	188	&	23124	$\pm$	31	&	-28	&	73	&	23145	$\pm$	27	&	-11	&	90	\\	
MSHT20an3lo\_as118~\cite{McGowan:2022nag}  	&	23101	$\pm$	47	&	-65	&	138	&	23138	$\pm$	29	&	-14	&	69	&	23154	$\pm$	26	&	-2	&	79	\\	
MSHT20qed\_an3lo~\cite{Cridge:2023ryv}                               	&	23101	$\pm$	52	&	-65	&	126	&	23137	$\pm$	31	&	-15	&	69	&	23143	$\pm$	28	&	-13	&	80	\\	\hline
PDF4LHC21\_40~\cite{PDF4LHCWorkingGroup:2022cjn}      	&	23122	$\pm$	83	&	-44	&	80	&	23133	$\pm$	33	&	-19	&	62	&	23143	$\pm$	27	&	-13	&	77	\\	\hline
\end{tabular}
\caption{
Values of \seff (in units of $10^{-5}$) before and after profiling with  the CMS 13 TeV  $A_4$ distribution (63 points) for 19 PDF sets.  
Also shown are the values extracted by including the CMS 13 TeV $W$-decay lepton asymmetry and the CMS W/Z cross section ratios (total of 83 points)  in the profiling. 
The  total uncertainties include contributions from statistical, experimental systematic, theoretical and PDF sources.  The column marked "Diff. from  \textsc{CT18Z}"  is the difference from  \seff extracted with the 
\textsc{CT18Znnlo} PDF set.
Note, the $\chi^2$ values before profiling do not include PDF errors, while the $\chi^2$ values after profiling include PDF errors.
 }
\label{Table1}
\end{center}
\vspace{-0.1in}
\end{table}

\section{\texorpdfstring{Extended Profiling with $W$ Charge Asymmetry and $W/Z$ Cross Section Ratios}{Extended Profiling with $W$ Charge Asymmetry and $W/Z$ Cross Section Ratios}}
After profiling, the dominant PDF uncertainty is largely reduced, but small differences between modern PDF sets are still observed. As shown in \cite{Bodek:2015ljm}, including the $W$-boson charge asymmetry measurements (\Wasym) in the profiling provides additional constraints on the $d/u$ ratio, leading to a further reduction of the PDF uncertainty in \seff. Moreover, some CT18 PDF sets lack sufficient constraints on the strange-quark distribution at high energy scales, resulting in systematically lower strange-quark densities compared to other PDFs and thereby predicting a larger effective weak mixing angle. Incorporating the $W/Z$ cross section ratios can provide further constraints on the strange-quark distribution, helping to reduce these discrepancies.

Therefore, we perform an updated analysis of the published CMS 13 TeV $A_4$ data (63 data points). 
We also include the recent CMS \Wasym measurement~\cite{CMS:2020cph} at 13 TeV (18 additional data points) 
and the CMS measurements of the $W$ and $Z$ fiducial cross-section ratios~\cite{CMS:2024myi} 
at 5.02 TeV ($12.505 \pm 0.037_{\rm stat} \pm 0.032_{\rm syst}$) 
and 13 TeV ($12.078 \pm 0.028_{\rm stat} \pm 0.032_{\rm syst}$), contributing 2 additional data points.
The results are shown in the columns of Table~\ref{Table1} and in Figure~\ref{Fig_2}.

  Extracted values of \seff from the 13 TeV CMS $A_4$ data  for different PDF sets are shown on the horizontal axis of  each of  the four panels in  Fig. \ref{Fig_2}.  The vertical  axis shows the values of $\chi ^2$ normalized to the number of degrees of freedom, $\rm{ndf}=N_{\rm data}-1$, where the subtraction accounts for the single free parameter \seff.  Shown are
the values before profiling (panel A, 63 data points).   After profiling with $A_4$ (panel B, 63 data  points), after profiling with $A_4$ and also with  $W$ decay lepton asymmetry (panel C,  81  data points), and after profiling with $A_4$ and the $W$ decay lepton asymmetry and also the W/Z cross section ratios (panel D, 83 data points).

As seen in Table~\ref{Table1} and Figure~\ref{Fig_2}, after $A_4$ profiling  the error in the extracted value of \seff with  the nominal \textsc{CT18Znnlo} PDF set is reduced from 0.00056 to 0.00032. By also  including the $W$ decay lepton asymmetry and the $W/Z$ cross section ratios the uncertainty for  the nominal \textsc{CT18Znnlo} is reduced to  0.00024.  

     \begin{wrapfigure}{l}{3.2in}
\includegraphics[width=3.2in,height=2.4in]{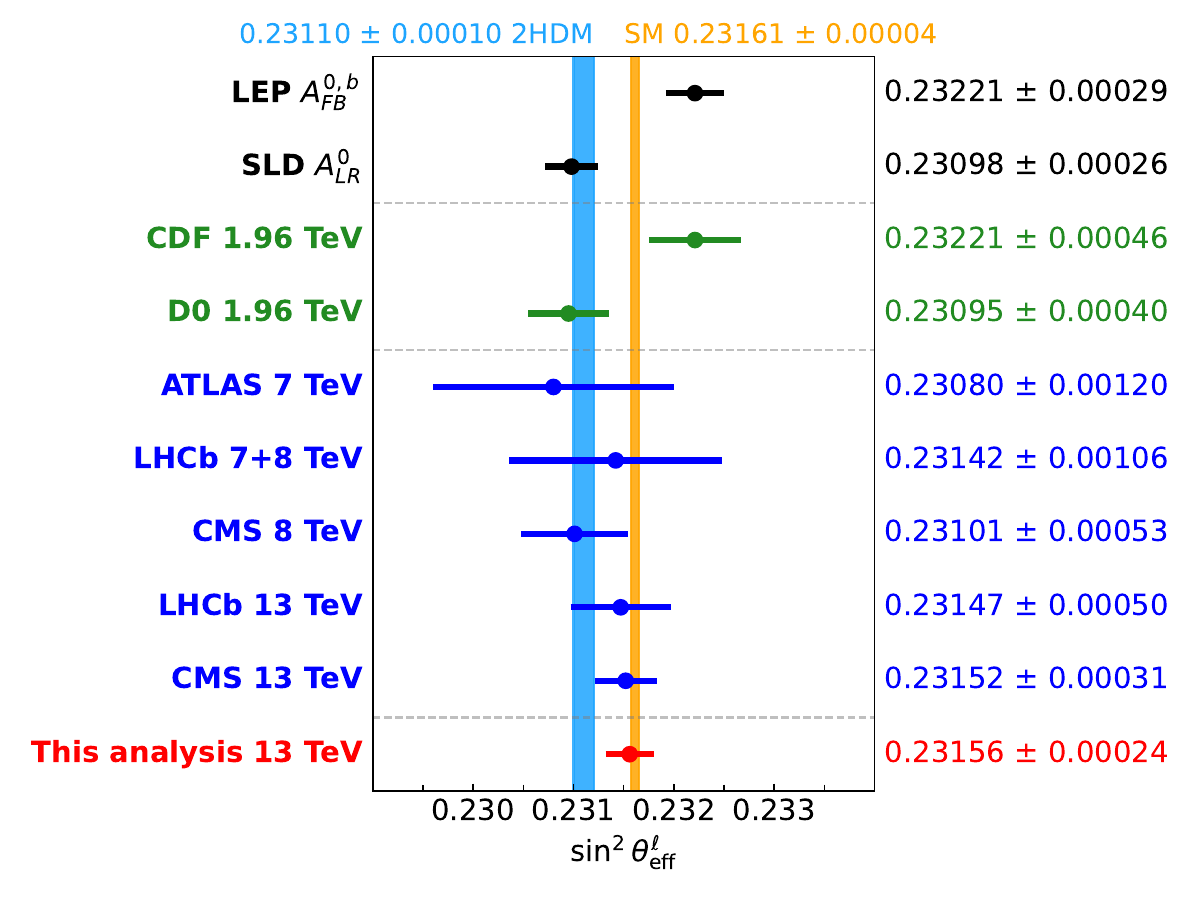}
\vspace{-0.4in}
\caption{ Comparison of \seff  extracted in this analysis (labeled "This analysis 13 TeV")
with previous measurements~\cite{ALEPH:2005ab,CDF:2016cei,D0:2017ekd,ATLAS:2015ihy,CMS:2018ktx,CMS:2024ony, LHCb:2015jyu,LHCb:2024ygc} 
and the prediction of the 2025 SM  global fit~\cite{ParticleDataGroup:2024cfk,PDG2025}.  Also shown is the prediction
of the Two Higgs Doublet Model~\cite{Biekotter:2022abc} corresponding to the   CDF $M_W$ value~\cite{CDF:2022hxs} 
(80.4335 $\pm$0.0094  GeV).
}
\label{Fig_7}
\vspace{-0.3in}
 \end{wrapfigure}

Including all three inputs in the profiling procedure results in a pronounced alignment among the different PDF sets.
Following profiling, the extracted central values from 18 of the 19 PDFs (including three of the four MSHT20 sets) are consistent with the \textsc{CT18Znnlo} determination at the level of one standard deviation.
The only exception is the profiled \textsc{MSHT20nnlo\_as118} set, whose central value lies 1.38 standard deviations below the \textsc{CT18Znnlo} result and is associated with a poor $\chi^2$.
By contrast, as reported in Table~\ref{Table1}, the profiled \textsc{MSHT20an3lo\_as118} set yields an identical value of \seff to that obtained with \textsc{CT18Znnlo}, together with an acceptable $\chi^2$.

 Figure~\ref{Fig_7} shows a comparison of \seff  extracted in this analysis  with the  nominal \textsc{CT18Znnlo}  PDF set  (Labeled "This analysis 13 TeV") 
 to previous measurements~\cite{ALEPH:2005ab,CDF:2016cei,D0:2017ekd,ATLAS:2015ihy,CMS:2018ktx,CMS:2024ony,LHCb:2015jyu,LHCb:2024ygc}. 
 Also shown are the predictions  of the 2025  SM global fit~\cite{ParticleDataGroup:2024cfk,PDG2025} and the prediction of the  Two Higgs Doublet Model~\cite{Biekotter:2022abc}
   assuming the CDF $M_W$ value~\cite{CDF:2022hxs} of 80.4335 $\pm$0.0094  GeV.

\section{Conclusion}
This proceeding reports on an improved extraction of the effective leptonic weak mixing angle, \seff, in which additional CMS measurements of the $W$-boson charge asymmetry and $W/Z$ cross-section ratios are incorporated to further constrain parton distribution function effects. The final result, $\seff = 0.23156 \pm 0.00024$, represents the most precise single measurement of this parameter to date, as presented in Ref.~\cite{Bodek:2025kxl}. A key outcome of this enhanced PDF profiling is the excellent consistency observed across 18 different PDF sets, with all results lying within one standard deviation of the nominal CT18ZNNLO value. The extracted value is compatible with the Standard Model expectation, $\seff = 0.23161 \pm 0.00004$.

Research supported by the U.S. Department of Energy under University of Rochester grant number DE-SC0008475 and by National Research Foundation of Korea (NRF) grants (RS-2024-00350406, RS-2008-NR007227).

{\footnotesize
\bibliographystyle{unsrt}
\bibliography{MixingAngle}
}
\end{document}